# Velocity Map Imaging the Scattering Plane of Gas Surface Collisions

David J. Hadden,[1] Thomas M. Messider,[1] Joseph G. Leng,[1] and Stuart J. Greaves[1,a)]

[1] *Institute of Chemical Sciences, Heriot-Watt University, Edinburgh, EH14 4AS, UK*

The ability of gas-surface dynamics studies to resolve the velocity distribution of the scattered species in the 2D scattering plane has been limited by technical capabilities and only a few different approaches have been explored in recent years. In comparison, gas-phase scattering studies have been transformed by the near ubiquitous use of velocity map imaging. We describe an innovative means of introducing a surface within the electric field of a typical velocity map imaging experiment. The retention of optimum velocity mapping conditions was demonstrated by measurements of iodomethane-$d_3$ photodissociation and SIMION calculations. To demonstrate the system's capabilities the velocity distributions of ammonia molecules scattered from a PTFE surface have been measured for multiple product rotational states.

It has been noted by several authors[1-5] that the study of gas-surface scattering could be revolutionized by using a combination of resonance enhance multiphoton ionization (REMPI)[6] and velocity map ion imaging (VMI).[7] REMPI-VMI allows the velocity distribution of quantum state selected products to be recorded; this approach has already been widely adopted in the gas-phase scattering community[8] as an vast improvement on previous techniques, *e.g.* traditional rotatable mass spectrometer approach[9], which do not detect the whole 2D scattering plane in a single measurement.

VMI uses the interpenetrating electric fields generated by annular electrodes to map ions with the same velocity, but created in different locations, onto the same spot on a position sensitive detector.[7, 10] Even small perturbations to these carefully created electric fields result in the loss of 'velocity mapping' conditions, so approaches to imaging the scattering from surfaces have endeavored to minimize these effects by either mounting the surface onto an electrode,[2-4] or outside the electrodes altogether.[1] These approaches mean that it is no longer possible to image the whole 2D scattering plane, and so multiple measurements are still required to obtain the total scattering distribution.

Detecting the velocity distribution of the whole 2D scattering plane in a single measurement requires the scattering plane to be parallel to the position sensitive detector. If the surface is mounted onto one of the electrodes[3, 4] then the scattering plane, which must contain the surface normal, is necessarily perpendicular to the

___________________________

a) Author to whom correspondence should be addressed. Electronic mail: s.j.greaves@hw.ac.uk.

detector preventing its direct measurement. Thus, either complex modeling or multiple time slicing measurements are required to deconvolute the velocity profile. Furthermore, only grazing angles of incidence are possible, as gas molecules have to pass between the electrodes to strike the surface.[3, 4]

Experiments that mount the surface outside the electrodes[1] are limited to having a large surface to laser distance; that means that the ionization region has to be particularly large to avoid preselecting molecules and biasing the measured distribution to a small range of scattering angles. This requires very shallow electric gradients that limit such techniques to only spatially imaging the scattered products and all velocity information has be inferred from images collected at individual time steps. In this paper, we present a novel adaptation to a standard VMI apparatus that overcomes these issues and provides a mechanism for directly imaging the scattering plane irrespective of the incident angle and should allow novel REMPI-VMI studies to be performed on numerous gas (dielectric)surface scattering systems.

The Surface-Scattering Velocity Map Imaging (SS-VMI) experimental set-up consisted of two differentially pumped chambers: a source chamber, that contained the molecular beam source; and a scattering chamber, which housed the laser focal region, the VMI electrodes (ion optics), the surface, and time-of-flight (TOF) region. The laser, TOF, and the molecular beam axes are mutually perpendicular (labeled x, y and z, respectively, see Figs. 1 and 2) and meet at the center of the scattering chamber.

The molecular beam source consisted of a pulsed solenoid valve (General Valve Corporation, series 9) with a 200 µs opening time, that was used to form a supersonic expansion of ammonia (2.5% in 4 Bar He), or iodomethane-$d_3$ (2.5% in 4 Bar Ar), with mean velocities of 1550 ms$^{-1}$ and 540 ms$^{-1}$, respectively. The molecular beam was collimated by a 1.01 mm skimmer (Beam Dynamics, model 1) mounted at the intersection between the two chambers, 33 mm from the front face of the nozzle and 176 mm from the laser axis.

A pulsed dye laser (Sirah Cobra-Stretch, DCM dye) pumped by a 20 Hz Nd:YAG laser (Continuum Surelite, SL I-20) was used to generate all required laser light. The laser beam had a diameter of 2 mm and was focused into the center of the scattering chamber by a 250 mm focal length lens. The focus created in the middle of the ion optics had a minimum beam waist of approximately 12 µm and a Rayleigh range of 1.4 mm. The generated ions were accelerated by the ion optics electric field along a 500 mm field free TOF region until they reach a position sensitive detector. The detector (Photek VID240 and GM-MCP-2) consisted of a pair of 40 mm diameter multi-channel plates (MCP) and a P-46 phosphor screen. Ion hits were recorded by a CCD camera (Basler scA780-54fm), with image acquisition and analysis performed by purpose written LabVIEW programs.

The ion optics (see Fig. 1) employed are based on those designed by Wrede *et al*.[11], which have seen widespread use in photodissociation experiments.[12-16] The design implemented in these studies (shown in fig 4 of ref. 12) comprised five electrodes: a cup shaped repeller, a conical extractor, two additional annular lenses and ground (Labelled R, E, L1, L2 and G respectively in Fig. 1). Careful optimization of the ion optics' voltages and fast pulsing of the second MCP (40 ns temporal width) allowed for dc-slice imaging, where the ion packets were elongated and only the central "slice" was acquired by pulsing the detector.[17] This allowed the exclusive imaging of only those products in the scattering plane defined by the molecular beam and the surface (*i.e.* the xz plane).

The surface studied in this work was PTFE (Polytetrafluoroethylene) with an exposed top face of 1 mm (along the TOF y-axis) by 12.7 mm (along the laser x-axis) that was normal to the molecular beam direction (z-axis, with the positive z direction defined as away from the surface). The PTFE surface was mounted in a PEEK (Polyether ether ketone) holder of the same cross-section, which extends approximately 50 mm down from the surface into a region outside the ion optics (shown in yellow in the lower half of Figs. 1 & 2). Two stainless steel scalpel blades (A.C.M. 18) were attached to the sides (parallel to the laser axis) of the PEEK holder and PTFE surface. These scalpel blades were employed as compensating electrodes to stabilize the electric field and maintain optimum velocity mapping conditions between the repeller and extractor. Scalpel blades were employed for these electrodes because of the dual benefits of their sharp edges: producing minimal electric field perturbation; and a narrow cross-section for scattering, which otherwise may mask scattering from the PTFE surface. The PEEK holder was mounted in the scattering chamber via an XYZ translator, which allowed the surface to be positioned relative to the TOF and laser axes, or removed from the electric field region altogether to perform comparative measurements using the "standard" VMI technique (referred to as VMI mode from hereon). All SS-VMI experiments were performed with the PTFE surface held 10 mm from, and aligned centrally to, the TOF and laser axes. Table I. shows the voltages applied to the ion optics and stabilizing electrodes in SS-VMI and VMI mode.

Table I. A summary of the voltages applied to each electrode, whether the surface was present within electric field of the ion optics.

|            | Repeller | Blade 1 | Blade 2 | Electrode | Lens 1 | Lens 2 |
|------------|----------|---------|---------|-----------|--------|--------|
| Surface    | 1700V    | 1628V   | 1598V   | 1505V     | 1273V  | 569V   |
| No Surface | 1700V    | -       | -       | 1500V     | 1273V  | 569V   |

Fig. 1 shows a 2D cross-section though the mid-point of the surface, in the yz plane (of the molecular beam and the TOF axes), modeled in SIMION (Scientific Instrument Services, Version 8.1), contours represent the electric field strength created by the ion optics. A perpendicular cross-section through the midpoint of the surface in the xz scattering plane, is shown in Fig. 2, including two calculated contour plots showing how the electric field changes with the presence of the surface, holder and stabilizing electrodes. Both figures show that the surface causes only a minimal perturbation to the curvature of the field.

To compare the velocity mapping conditions of the experiment in VMI and SS-VMI modes one color photodissociation of $CD_3I$ and subsequent ionization of $I(^2P_{1/2})$ atoms was performed at 310.6 nm, using a (2 + 1) REMPI transition (via the $5p^4(^3P_0) 6p^1\ ^2[1]°_{3/2}$ state).[18] This acted as a useful benchmark for velocity calibration. Fig. 3 shows the resultant images, with Fig. 3(a) taken in VMI mode (without the surface present) and Fig. 3(b) in SS-VMI mode. From a comparison of the images, which both show distinct rings, it is clear that introducing the surface had minimal effect on the velocity mapping conditions, with the radii of the rings varying by less than a pixel. This negligible difference is also emphasized by the small change of detector slice timing required to obtain the center of the iodine ion packet, a 5 ns change on a 16.24 μs flight time (0.03%).

Images of surface scattered $NH_3$ were recorded in three different rotational states: $J_K = 1_0$, $5_3$ and a combination of the $8_6/10_9$ levels in a single band; and are shown in Fig. 4. Ionization of incoming and scattered $NH_3$ molecules was by (2 + 1) REMPI via the $^1\tilde{B}(v_2 = 4) \leftarrow\ ^1\tilde{X}(v = 0)$ transition in the range 316.5 – 318.5 nm.[19] The rotational levels studied were isolated spectroscopic transitions that represent low, medium and high $J_K$ cases in the scattered signal. Each $NH_3$ scattering image contains two features: signal from the incident molecular beam (lower in Fig. 4) and signal from surface scattering (upper in Fig. 4). As scattered molecules must make a 20 mm round trip from the detection region to the surface and back again, scattering signal is expected to appear at a minimum time of 13 μs after the initial molecular beam appearance; the images in Fig. 4 are taken at a 90 μs delay.

It should be noted that the slight perturbation of the electric field, predicted by the SIMION calculations (Fig. 2), is far more significant for ions whose lab frame velocity is directed towards the surface (negative z), while ions with a lab frame velocity away from the surface (positive z) will experience minimal perturbation of the electric field. For the $CD_3I$ images in Fig. 3b both the upper and lower crescents of the rings have negative z velocities in the lab frame, as the photodissociation induced velocity of the iodine is less than the molecular beam velocity. However, the $NH_3$ scattered signal only possess positive z velocity taking these ions away from the surface to an area where the velocity mapping field is unperturbed by the presence of the surface.

Speed distributions of the scattered $NH_3$ are obtained by radially integrating the signal in a 90° wedge centered on the positive z direction from zero lab frame velocity. Radial integration of 90° wedges perpendicular to the molecular beam (*i.e.* along the x-axis), from zero lab frame velocity, was used to obtain and subtract the background signal caused by $NH_3$ scattering from the chamber walls and electrodes. This background is visible in Fig 4. close to zero lab frame velocity. The radial signal is then converted to speed by a calibration factor (17.92 ms$^{-1}$ pixel$^{-1}$ for $NH_3$) determined from the $CD_3I$ images. Normalized speed profiles of the scattering signal (from Fig. 4) are shown in Fig. 5. These speed profiles show that there is a large exchange of translational energy during the collision between $NH_3$ and surface; by fitting the speed profile to a Maxwell-Boltzmann distribution a mean $NH_3$ speed[20] can be obtained (694 ms$^{-1}$ for $J_K = 1_0$) that is significantly slower than the incoming molecular beam velocity. We note that the loss in translational energy does not follow a trend dependent on J, as the mean speed increases from $J_K = 1_0$ to $5_3$ (821 ms$^{-1}$) and then decreases again from $5_3$ to $8_6/10_9$ (799 ms$^{-1}$).

The angular distribution of scattered signal about the zero lab frame velocity in Fig. 4 show a narrow range with all signal in an approximately 75° wedge centered around the z axis. The angular distribution of signal shows only a slight change across the three rotational levels studied, with $J_K = 1_0$ showing a narrower spread by approximately 15°. It should be noted that the range of measurable scattering angles is limited by our detection geometry; this limit is close to an 80° region, centered on the positive z direction from the zero lab frame velocity, calculated using the Raleigh range of laser focus and the surface to laser distance.

In conclusion, we have demonstrated the viability of SS-VMI as a means of directly detecting the 2D velocity distribution in the scattering plane of a gas-surface collision, which can lead to more rigorous studies of scattering dynamics. The calculations of electric field contours and the experimental iodomethane-d$_3$ photodissociation images have shown the feasibility of using stabilizing electrodes to maintain velocity mapping conditions while introducing a dielectric object (in this case the PTFE surface) inside the VMI ion optics. Ammonia scattering results have highlighted the quality of the velocity distributions that can be obtained with these experiments. Although the range of detectable scattering angles currently limits the technique, it is expected that an alternative ion optics design will allow a reduction of the surface to laser distance (while maintaining velocity mapping conditions). Coupled with greater laser Rayleigh range, this will minimize the level of background signal, improve velocity resolution and allow a greater than 150° range of measurable angles.

This research was possible thanks to an EPSRC Career Acceleration Fellowship for S.J.G. (EPSRC EP/J002534/2). T.M.M. is grateful to Heriot-Watt University for a PhD studentship. We would like to thank Cassandra Rusher for experimental assistance; Dr Matt Costen and Prof. Ken McKendrick for their helpful discussions and loan of experimental equipment.

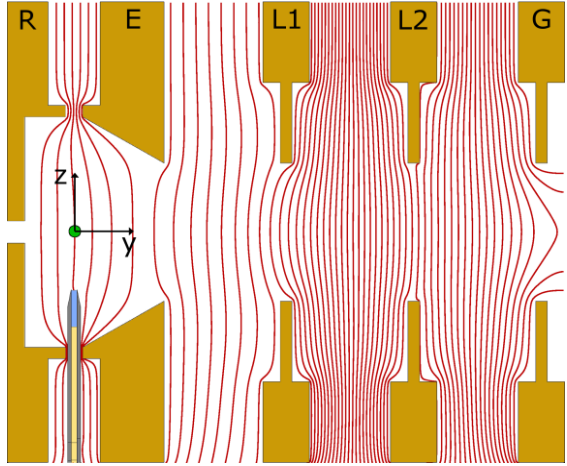

FIG. 1. A 2D cross section of the ion optics (orange), the surface and its mounting in the yz plane. The repeller, extractor, two lenses and ground are labelled R, E, L1, L2 and G, respectively. The PTFE section is shown in blue, PEEK holder in yellow and the scalpel blades on either side shown in gray. The laser position is represented by the green dot. The red contours represent the SIMION calculated electric field and are spaced at 30 V intervals.

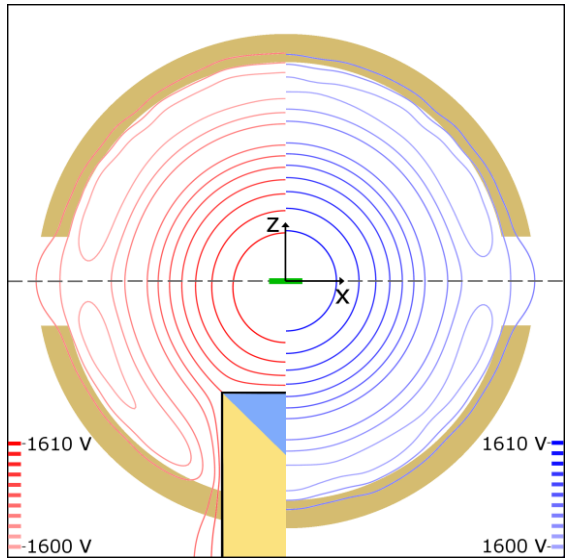

FIG. 2. 2D cross section through the ion optics (orange) in the xz plane, taken at the midpoint through the surface. The left side with red contours represents the imaging field in the presence of the PTFE surface section (light blue) and its mounting (yellow). The right half with the blue contours shows the field without the surface present. The green rectangle represents the ionization region.

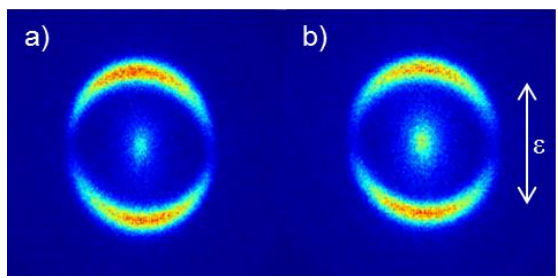

FIG. 3. a) Velocity Map Image and b) Surface Scattering Velocity Map Image of iodine cations formed by one color photodissociation of iodomethane-d3 as described in the text. The electric field polarization of the dissociation laser is noted by the double-ended white arrow in b).

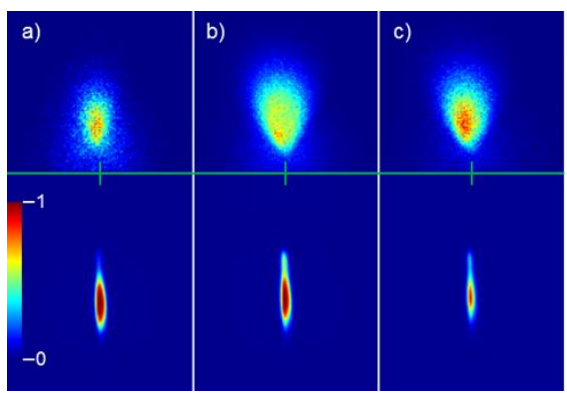

FIG. 4. Surface Scattering Velocity Map Images of ammonia cations formed from the $J_K$ = a) $1_0$, b) $5_3$ and c) $8_6/10_9$ rotational levels of the $\tilde{X}$ state. The intersection of the green lines corresponds to zero lab frame velocity; below the horizontal line the molecular beam velocity distribution is shown; above this line shows surface scattered ammonia cation signal with velocities opposing the molecular beam (positive z).

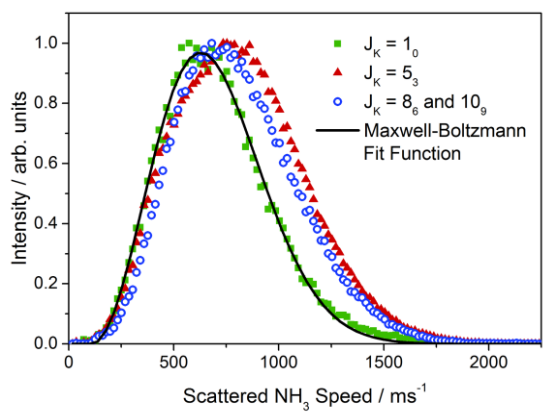

FIG. 5. Scattered ammonia speed distributions derived from Fig. 4, measured for the $J_K$ = $1_0$, $5_3$ and $8_6/10_9$ rotational levels of the ammonia $\tilde{X}$ state. A Maxwell-Boltzmann distribution is fit to the $J_K$ = $1_0$ signal, with a mean ammonia velocity of 694 ms$^{-1}$